\documentclass[pre,preprint,showpacs,showkeys,preprintnumbers,amsmath,amssymb]{revtex4}
\usepackage{graphicx}
\usepackage{dcolumn}
\usepackage{bm}

\begin{document}
\newcommand{\re}{\mathrm{e}}            
\title{Random walks on finite lattice
tubes}

\author{B.I. Henry}
\email{b.henry@unsw.edu.au}
\affiliation{Department of Applied Mathematics, School
of Mathematics, University of New South Wales,
Sydney NSW 2052, Australia\\}

\author{M.T. Batchelor}
\email{murrayb@maths.anu.edu.au}
\affiliation{
Department of Theoretical Physics,
Research School of Physical Sciences and Engineering,
and Centre for Mathematics and its Applications,
Mathematical Sciences Institute,
Australian National University  
Canberra ACT 0200, Australia\\}

\date{today}

\begin{abstract}
Exact results are obtained for random walks
on finite lattice tubes with a single source and absorbing
lattice sites at the ends.
Explicit formulae are derived for the absorption probabilities
at the ends
and for the expectations that a random
walk will visit a particular lattice site before being absorbed.
Results are obtained for lattice tubes of arbitrary size
and each of the regular lattice types;
square, triangular and honeycomb. The results include an adjustable parameter
 to model the effects of strain, such as surface curvature,
on the surface diffusion.
Results for the triangular lattice tubes and the honeycomb lattice tubes
 model diffusion of adatoms
 on single walled zig-zag carbon nano-tubes 
with open ends.

\end{abstract}


\maketitle
\section{Introduction}

The problem of random walks on finite lattices
is fundamental to the
theory of stochastic processes \cite{ML79} and has numerous applications
including
potential theory \cite{D53},
electrical networks \cite{DS84}, atomic
surface diffusion \cite{BCMO99} and diffusion on biological membranes
\cite{HPAB99}.
 A classic problem in this area, which was  posed
by Courant  \textit{et al.} \cite{C28} in 1928, concerns
random walks on finite planar lattices with a single source
and absorbing boundaries.
The exact solution for this problem
on the square lattice was derived in 1940
\cite{MW40}. The exact solution
on the triangular lattice
was only obtained recently \cite{BH02c,BH02} after having been considered
intractable \cite{KM63}.
Other variants of the problem on the square lattice
have also been solved exactly \cite{KM63,M94,FZ01}.

In this paper we consider random walks from a single source
on a finite lattice which incorporates periodic
boundary conditions in one direction but
absorbing boundary conditions in
the other. Diffusion thus occurs on a surface
with the topology of a
lattice tube with absorbing sites at the ends.
 We present explicit results for
each of the three regular lattice types; square, triangular
and honeycomb.
Our results include a bias parameter that can be adjusted
away from unity to model different random walk probabilities in
the cyclic
direction around the tube compared with
the axial direction along the tube.
This parameter can be adjusted to model the effect of surface curvature,
and other types of strain,
on the surface diffusion.

Our derivations generalize
the approach developed by McCrea and Whipple
\cite{MW40} for planar square lattice random walks with absorbing boundaries.
This is straightforward in the case of the square lattice tube however
special
care has to be exercised in an appropriate choice
of co-ordinates in the case of
the triangular lattice tube and the honeycomb lattice tube.

There are two important motivations for our study.
One of the motivations is to add to
the rather small class of exactly solvable random walk lattice problems
with absorbing boundaries, since it is still the case that:
 ``Explicit solutions are known in only
 a few cases'' \cite{F66}.
The second motivation is
that
 the mathematics
of discrete lattice diffusion problems may
find applications in
the recently realized
laboratory
assembly of lattice nanostructures (see for example, \cite{LJLLWXLTZZ02}).
For example our problem of random walks
on the triangular lattice tube and the honeycomb lattice tube
represent models for
adatom diffusion on
single walled zig-zag carbon nano-tubes \cite{O98} with open ends --
the random walks on the triangular lattice
tube model adatom diffusion across carbon-carbon bonds
and the random walks on the honeycomb lattice
tube model diffusion along the carbon-carbon bonds.
The diffusion of carbon adatoms along the carbon-carbon bonds of a
carbon nanotube plays a vital role in stabilizing and maintaining the
open edge growth of nanotubes \cite{KO01,LSK02}.
Our exact results complement related results for diffusion on carbon
nano-tubes based on i) enumeration of random walks up to a set length
\cite{C00} and ii)
microcanonical molecular
dynamics simulations \cite{SG01}.

The remainder of the paper is divided into separate sections
for square lattice tubes, triangular lattice tubes,
honeycomb lattice tubes, and a section  containing
an example and discussion.

\section{Square lattice tubes}
Consider the standard square lattice co-ordinates
 $(p,q)$ representing the intersections
of equidistant vertical straight lines and
equidistant horizontal straight lines.
The expectation that a random walk starting from a site $(a,b)$ visits
a site $(p,q)$, distinct from $(a,b)$, before being
absorbed at a finite boundary site is given by the homogeneous
partial difference
equation
\begin{eqnarray}
F(p,q)&=&\frac{1}{2+2\eta}\left[F(p+1,q)+F(p-1,q)\right.\nonumber\\
& &\qquad\left.+\eta F(p,q+1)+ \eta F(p,q-1)\right].\label{SH}
\end{eqnarray}
The parameter $\eta$ allows for different probabilities for
walks around the tube compared with walks along the tube.
To accommodate the source term at $(a,b)$ we construct separate solutions;
$F_I(p,q)$ for $q\le b$ and $F_{II}(p,q)$ for $q\ge b$.
The expectation that a random walk starting from
$(a,b)$ visits a site $(p,b)$,
not necessarily distinct from  $(a,b)$, before being
absorbed at a finite boundary is then given by
the inhomogeneous partial difference equation
\begin{eqnarray}
F_I(p,b)&=&\delta_{p,a}+\frac{1}{2+2\eta}\left[F_I(p+1,b)+F_I(p-1,b)\right.\nonumber\\
& &\qquad\left.+\eta F_{I}(p,b-1)
+\eta F_{II}(p,b+1)\right].\label{SI}
\end{eqnarray}
The above difference equations are to be solved with
periodic boundary conditions in the $p$ co-ordinates,
\begin{equation}
F(p,q)=F(p+m+1,q), \label{SBC1}
\end{equation}
absorbing boundary conditions in the $q$ co-ordinates,
\begin{eqnarray}
F_I(p,0)=0,\label{SBC2}\\
F_{II}(p,n+1)=0\label{SBC3},
\end{eqnarray}
and matching conditions at $q=b$, ie., 
\begin{equation}
F_I(p,b)=F_{II}(p,b). \label{SM}
\end{equation}

The method of solving inhomogeneous linear partial difference
boundary value problems
as above consists of two parts \cite{MW40,KM63,F66,BH02}.
First obtain the general separation of variables solution to the
 homogeneous problem, then find an appropriate linear combination
of such solutions to satisfy the boundary conditions
and the inhomogeneous problem.
A major difficulty in these problems can be
the identification of a lattice co-ordinate system with
a separation of variables solution that can be matched with the
 boundary conditions \cite{BH02}.

The homogeneous field equations
for the square lattice, Eq.(\ref{SH}), admit the separable solution
\begin{equation}
F(p,q)=P(p)Q(q)
\end{equation}
where
\begin{eqnarray}
P(p+1)+(\lambda-(2+2\eta))P(p)+ P(p-1)&=&0,\\
Q(q+1)-\frac{\lambda}{\eta}Q(q)+Q(q-1)&=&0,
\end{eqnarray}
and $\lambda$ is the separation constant.
These separated equations have general solutions
\begin{eqnarray}
P(p)&=&A\mu^p+B\mu^{-p},\label{SPH}\\
Q(q)&=&C\nu^q+D\nu^{-q}\label{SQH},
\end{eqnarray}
where
\begin{eqnarray}
\mu&=&\frac{2+2\eta-\lambda}{2}+\frac{\sqrt{(2+2\eta-\lambda)^2-4}}{2},\\
\nu&=&\frac{\lambda}{2\eta}+\frac{\sqrt{\frac{\lambda^2}{\eta^2}-4}}{2}.
\end{eqnarray}
If $\lambda=2\eta$ then the solutions
are no longer provided by Eqs. (\ref{SPH}) and (\ref{SQH}).
In this case we have the solutions
\begin{eqnarray}
P(p)&=&\hat A +\hat B p\label{SPh}\\
Q(q)&=&\hat C+\hat D q\label{SQh}.
\end{eqnarray}
The special solutions in Eqs.(\ref{SPh}) and (\ref{SQh}) do not appear in
the planar lattice problems because they cannot satisfy absorbing boundary
conditions in both $p$ and $q$ co-ordinates.

With suitable linear combinations of the above solutions, Eqs.(\ref{SPH}),(\ref{SQH}),(\ref{SPh}),(\ref{SQh}),
we find that the general solutions to the homogeneous problem that also satisfy
the boundary conditions, Eqs.(\ref{SBC1}),(\ref{SBC2}),
 and the matching condition, Eq.(\ref{SM}),
 can be written in the form
\begin{eqnarray}
F_I(p,q)&=&cq(b-n-1)+\sum_{k=1}^m (a_k e^{i\alpha_k p}+b_k e^{-i\alpha_k p})\nonumber\\
& &\quad\times \sinh(\beta_k q) \sinh[\beta_k(b-n-1)]\\
F_{II}(p,q)&=&cb(q-n-1))+\sum_{k=1}^m (a_k e^{i\alpha_k p}+b_k e^{-i\alpha_k p})\nonumber\\
& &\quad\times \sinh[\beta_k(q-n-1)]\sinh(\beta_k b)
\end{eqnarray}
where
\begin{equation}
\alpha_k=\frac{2\pi k}{m+1} \label{alp}
\end{equation}
and
\begin{equation}
2+2\eta=2\eta\cosh\beta_k+2\cos\alpha_k \label{bet}
\end{equation}

The constants $c$, $a_k$ and $b_k$ are now determined by the requirement that
the solutions satisfy the inhomogeneous equation, Eq.(\ref{SI}). This step
is facilitated using the identity in Eq.(\ref{bet}) together with the identity
in Eq.(\ref{A1}) and
the Kronecker delta identity
\begin{equation}
\sum_{k=0}^m e^{2\pi i(p-a)k/(m+1)}=(m+1)\delta_{p,a}
\end{equation}
re-arranged as
\begin{eqnarray}
\delta_{p,a}=\frac{1}{m+1}+\frac{1}{2(m+1)}\left(\sum_{k=1}^m
e^{i\alpha_k p}e^{-i\alpha_k a}\right.&&\nonumber\\
\quad \left.+\sum_{k=1}^m
e^{-i\alpha_k p}e^{i\alpha_k a}\right).&&\label{Id}
\end{eqnarray}
We thereby
obtain the following solutions for the expectations that a random walk
will visit a site $(p,q)$ before being absorbed at an end site:
\begin{eqnarray}
F_I(p,q)=&&\frac{(2+2\eta)q(n+1-b)}{\eta(n+1)(m+1)}\nonumber\\
&&+\frac{2+2\eta}{\eta(m+1)}\sum_{k=1}^m\cos[\alpha_k(p-a)]\nonumber\\
&& \times
\frac{\sinh[\beta_k(n+1-b)]\sinh(\beta_k q)}{\sinh(\beta_k)\sinh[\beta_k(n+1)]},\label{SS1}
\end{eqnarray}
\begin{eqnarray}
F_{II}(p,q)=&&\frac{(2+2\eta)b(n+1-q)}{\eta(n+1)(m+1)}\nonumber\\
&&+\frac{2+2\eta}{\eta(m+1)}\sum_{k=1}^m\cos[\alpha_k(p-a)]\nonumber\\
&& \times\frac{\sinh[\beta_k(n+1-q)]\sinh(\beta_k b)}{\sinh(\beta_k)\sinh[\beta_k(n+1)]},\label{SS2}
\end{eqnarray}
with $\beta_k$ dependent on $k$ through Eqs.(\ref{alp}),(\ref{bet}).

The  absorption probabilities $G(p,q)$ are readily evaluated from
\begin{eqnarray}
G(p,0)&=&\frac{\eta}{2+2\eta}F_I(p,1),\\
G(p,n+1)&=&\frac{\eta}{2+2\eta}F_{II}(p,n).
\end{eqnarray}
It
is a simple matter to show that
$$
\sum_{p=0}^m G(p,0)=\left(\frac{n+1-b}{n+1}\right)
$$
and
$$
\sum_{p=0}^m G(p,n+1)=\left(\frac{b}{n+1}\right)
$$
so that
$$
\sum_{p=0}^m\left[G(p,0)+G(p,n+1)\right]=1.
$$

\section{Triangular lattice tubes}
To obtain the solution for triangular lattice tubes we consider the
lattice co-ordinates $(p,q)$
shown in Figure 1. This co-ordinate system describes
two independent triangular lattice systems, only one of which
can be accessed by a random walk from a single point source.  In Figure 1 
the sites that are accessible from the source site at $(a,b)$
are indicated by filled circles. The nearest neighbour
sites to the source are highlighted by open circles in this figure.
The lattice sites that are not accessible from the source are referred
to as the zero mesh \cite{KM63}.
\begin{figure}[h!]
\vspace{120mm}
\includegraphics{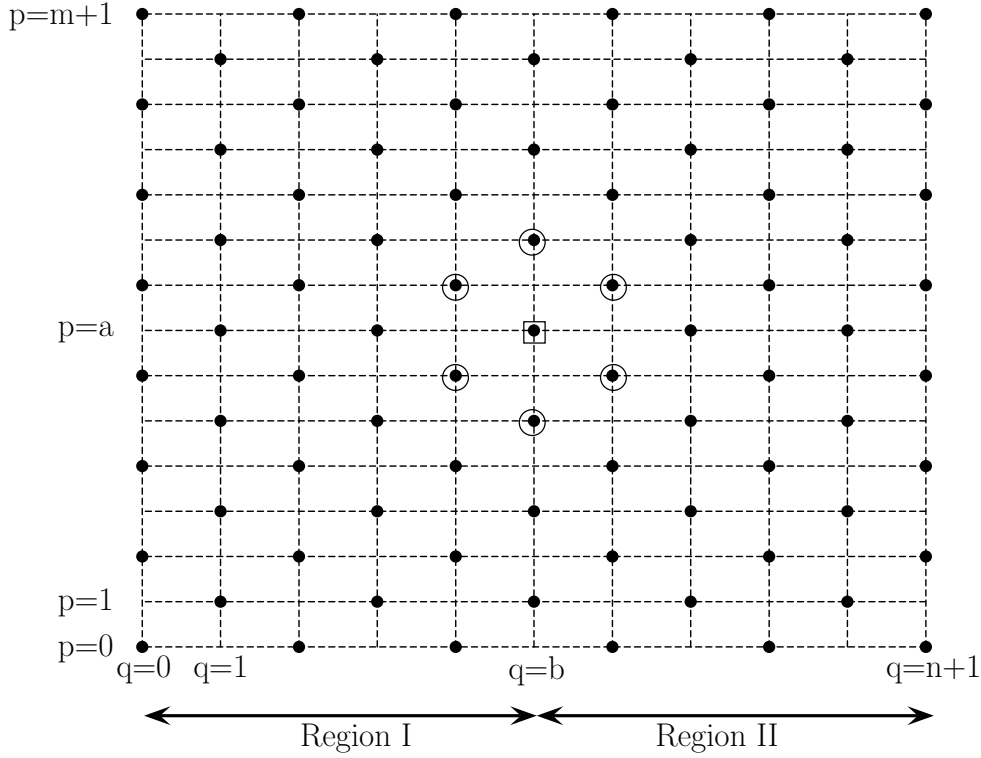}
\caption{Triangular lattice with $(p,q)$ co-ordinates
as used in section III.}
\end{figure}
The same co-ordinate system has been used to find an approximate
solution to the planar triangular lattice problem with
absorbing boundaries \cite{KM63} however a different 
zig-zag co-ordinate system was required to find the exact solution
\cite{BH02}. In the case of the planar triangular lattice
problem the co-ordinate system in Figure 1 allows the leakage of random
walks from sites at $p=1$ and $p=m$ to sites at $p=-1$ and $p=m+2$
respectively,
which are outside the absorbing boundaries at $p=0$ and $p=m+1$.
In the triangular lattice tube this
leakage is prevented because the $p$ co-ordinate is  cyclic.
Note that $m+1$ must be an even integer to permit
periodic boundaries in the $p$ direction.

With the co-ordinate system shown in Figure 1
 the
problem on the triangular lattice tube
is described by the homogeneous equation, for $(p,q)\ne (a,b)$,
\begin{eqnarray}
F(p,q)=\frac{1}{2+4\eta}\left[F(p+2,q)+F(p-2,q)\right.&&\nonumber\\
+\eta F(p+1,q+1)+\eta F(p+1,q-1)&&\nonumber\\
\left.+\eta F(p-1,q+1)+\eta F(p-1,q-1)\right],&&\label{TH}
\end{eqnarray}
the inhomogeneous equation, for $q=b$,
\begin{eqnarray}
F_I(p,b)=\delta_{p,a}
+\frac{1}{2+4\eta}\left[F_I(p+2,b)+F_I(p-2,b)\right.&&\nonumber\\
+\eta F_{II}(p+1,b+1)+\eta F_I(p+1,b-1)&&\nonumber\\
\left.+ \eta F_{II}(p-1,b+1)+\eta F_I(p-1,b-1)\right],&&\label{TI}
\end{eqnarray}
and the boundary conditions, Eqs.(\ref{SBC1}),(\ref{SBC2}),(\ref{SM}).

The homogeneous equation separates as
\begin{eqnarray}
P(p+2)+\lambda P(p+1)-(2+4\eta)P(p)& &\nonumber\\
+ \lambda P(p-1) +P(p-2)&=&0,\\
Q(q+1)-\frac{\lambda}{\eta} Q(q)+Q(q-1)&=&0,
\end{eqnarray}
with solutions 
\begin{eqnarray}
P(p)&=&Ae^{i\alpha p}+B^{-i\alpha p}+Ce^{i\beta p}+De^{-i\beta p},\\
Q(q)&=&Ee^{\gamma q}+Fe^{-\gamma q},
\end{eqnarray}
where
\begin{eqnarray}
\alpha&=&\cos^{-1}\left(-\frac{\lambda}{4}+\frac{1}{4}\sqrt{\lambda^2+16+16\eta}\right),\\
\beta&=&\cos^{-1}\left(-\frac{\lambda}{4}-\frac{1}{4}\sqrt{\lambda^2+16+16\eta}\right),\\
\gamma&=&\cosh^{-1}(\frac{\lambda}{2\eta}).
\end{eqnarray}
The general solution to the
 homogeneous problem on the triangular lattice
that satisfies all the boundary conditions,
 Eqs.(\ref{SBC1}),(\ref{SBC2}),(\ref{SM}), can be written as
\begin{eqnarray}
F_I(p,q)&=&cq(b-n-1)\nonumber\\
& &+d q(b-n-1)\cos[\pi(p-a)]\cos[\pi(q-b)]\nonumber\\
& &+{\sum_{k=1}^m}^\prime (a_k e^{i\alpha_k p}+b_k e^{-i\alpha_k p})\nonumber\\
& &\quad\times \sinh(\gamma_k q) \sinh[\gamma_k(b-n-1)],\label{Th1}\\
F_{II}(p,q)&=&cb(q-n-1)\nonumber\\
& &+d b(q-n-1)\cos[\pi(p-a)]\cos[\pi(q-b)]\nonumber\\
& &+{\sum_{k=1}^m}^\prime (a_k e^{i\alpha_k p}+b_k e^{-i\alpha_k p})\nonumber\\
& &\quad\times \sinh[\gamma_k(q-n-1)]\sinh(\gamma_k b),\label{Th2}
\end{eqnarray}
where the prime on the sum has been used to
indicate that the sum
does not include the value $k=(m+1)/2$ and
\begin{equation}
\alpha_k=\frac{2\pi k}{m+1},\label{Talp}
\end{equation}
with
\begin{equation}
2\eta\cosh\gamma_k\cos\alpha_k=1+2\eta-\cos2\alpha_k.\label{Tgam}
\end{equation}
The homogeneous solutions of the form
\begin{equation}
(\hat A +\hat B q) \cos[\pi(p-a)]\cos[\pi(q-b)]
\end{equation}
in Eqs.(\ref{Th1}),(\ref{Th2}),
are important in two fundamental ways.
First they replace the null solutions at $k=(m+1)/2$ in the
 representation 
$$
(a_k e^{i\alpha_k p}+b_k e^{-i\alpha_k p})\sinh(\gamma_k q) \sinh[\gamma_k(b-n-1)]
$$
and secondly they allow the appropriate zero mesh solution
for lattice co-ordinates that cannot be accessed by
a
source at $(a,b)$.

The constants $c,d,a_k,b_k$ are 
found by substituting the homogeneous solutions, 
Eqs.(\ref{Th1}),(\ref{Th2}), 
 into the inhomogeneous equation, Eq.(\ref{TI}).
Using the identity in Eq.(\ref{Tgam}) together with the idendity in
Eq.(\ref{A1}) we first obtain the intermediate
result
\begin{eqnarray}
-(2+4\eta)\delta_{p,a}=2\eta c(n+1)+2\eta d(n+1)\cos[\pi(p-a)]&&\nonumber\\
\quad +2\eta{\sum_{k=1}^m}^\prime (a_ke^{i\alpha_k p}+b_ke^{-i\alpha_k p})
\cos\alpha_k\sinh\gamma_k\sinh[\gamma_k(n+1)]&&.
\end{eqnarray}
The final result is then found by expanding the Kronecker delta as
\begin{eqnarray}
\delta_{p,a}=\frac{1}{m+1}+
\frac{1}{m+1}\cos[\pi(p-a)]\nonumber\\
+\frac{1}{2(m+1)}\left({\sum_{k=1}^m}^\prime
e^{i\alpha_k p}e^{-i\alpha_k a}+
e^{-i\alpha_k p}e^{i\alpha_k a}\right).&&\label{Id2}
\end{eqnarray}
Thus we obtain
\begin{eqnarray}
F_I(p,q)=&&\frac{(1+2\eta)q(n+1-b)}{\eta(n+1)(m+1)}\nonumber\\
&&\times (1+\cos[\pi(p-a)]\cos[\pi(q-b)])\nonumber\\
&&+\frac{1+2\eta}{\eta(m+1)}{\sum_{k=1}^m}^\prime\cos[\alpha_k(p-a)]
\nonumber\\
&& \times
\frac{\sinh[\gamma_k(n+1-b)]\sinh(\gamma_k q)}{\cos\alpha_k\sinh\gamma_k\sinh[\gamma_k(n+1)]},\label{T1}
\end{eqnarray}
\begin{eqnarray}
F_{II}(p,q)=&&\frac{(1+2\eta)b(n+1-q)}{\eta(n+1)(m+1)}\nonumber\\
&&\times (1+\cos[\pi(p-a)]\cos[\pi(q-b)])\nonumber\\
&&+\frac{1+2\eta}{\eta(m+1)}{\sum_{k=1}^m}^\prime \cos[\alpha_k(p-a)]
\nonumber\\
&& \times\frac{\sinh[\gamma_k(n+1-q)]\sinh(\gamma_k b)}{\cos\alpha_k\sinh\gamma_k\sinh[\gamma_k(n+1)]},\label{T2}
\end{eqnarray}
where 
$\alpha_k$ and $\gamma_k$ depend on $k$ through Eqs.(\ref{Talp}),(\ref{Tgam}).
Our results in Eqs.(\ref{T1}),(\ref{T2})
hold for arbitrary $m$ except
$m+1 = 0 (\mbox{mod}\quad 4)$
where singularities occur for
$k=\frac{m+1}{4},\frac{3(m+1)}{4}$.
The  absorption probabilities $G(p,q)$ are given by
\begin{eqnarray}
G(p,0)=\frac{\eta}{2+4\eta}\left(F_I(p+1,1)+F_I(p-1,1)\right),&&\\
G(p,n+1)=\frac{\eta}{2+4\eta}\left(F_{II}(p+1,n)+F_{II}(p-1,n)\right),&&.
\end{eqnarray}
The absorption probabilities at the ends of the tube are
thus 
$$
\sum_{p=0}^m G(p,0)=\left(\frac{n+1-b}{n+1}\right)
$$
and
$$
\sum_{p=0}^m G(p,n+1)=\left(\frac{b}{n+1}\right)
$$
with
$$
\sum_{p=0}^m\left[G(p,0)+G(p,n+1)\right]=1.
$$

\section{Honeycomb lattice tubes}
\begin{figure}[h!]
\vspace{120mm}
\includegraphics{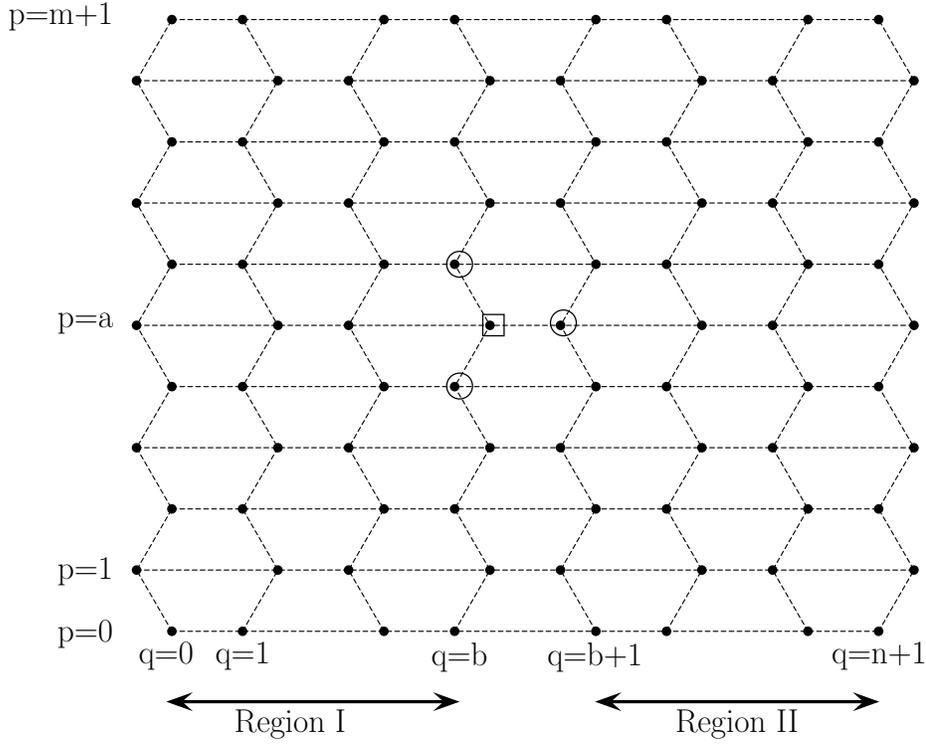}
\caption{Honeycomb lattice with $(p,q)$ co-ordinates
as used in section IV.}
\end{figure}
Here we consider 
a co-ordinate system in which we label the vertices
 of the honeycomb lattice by the intersection points $(p,q)$
 of horizontal straight lines
$p=0,1,2,\ldots , m+1$ and vertical zig-zag lines $q=0,1,2,\ldots , n+1$
(see Figure 2). The expectation
that a random walk 
visits an interior site $(p,q)$, distinct from the starting site
$(a,b)$, is given by the coupled homogeneous difference equations:
\begin{eqnarray}
F(p,q)=\frac{1}{2+\eta}&&\left[\eta \hat F(p,q+1)+ \hat F(p+1,q)\right.\nonumber\\
&&\qquad \left. +
 \hat F(p-1,q)\right],\label{H1}\\ 
\hat F(p,q)=\frac{1}{2+\eta}&&\left[\eta F(p,q-1)+ F(p+1,q)\right.\nonumber\\
&&\qquad \left. + F(p-1,q)\right].\label{H2}
\end{eqnarray}
Here $F(p,q)$ is the expectation at sites $(p,q)$ with
nearest neighbours on the right at $(p,q+1)$ and
$\hat F(p,q)$ is the expectation at sites $(p,q)$ with nearest neighbours
on the left at $(p,q-1)$. We will refer to these distinct symmetry sites
as $\vdash$ sites and $\dashv$ sites respectively.
The appeal of this particular choice of $(p,q)$
co-ordinates is that the difference equations for the
distinct symmetry sites can be decoupled into 
separable equations
for each. Indeed both $F(p,q)$ and $\hat F(p,q)$ satisfy the same
homogeneous
 triangular lattice equation as Eq.(\ref{TH}) except near the
absorbing boundaries.
\begin{figure}[h!]
\vspace{110mm}
\includegraphics{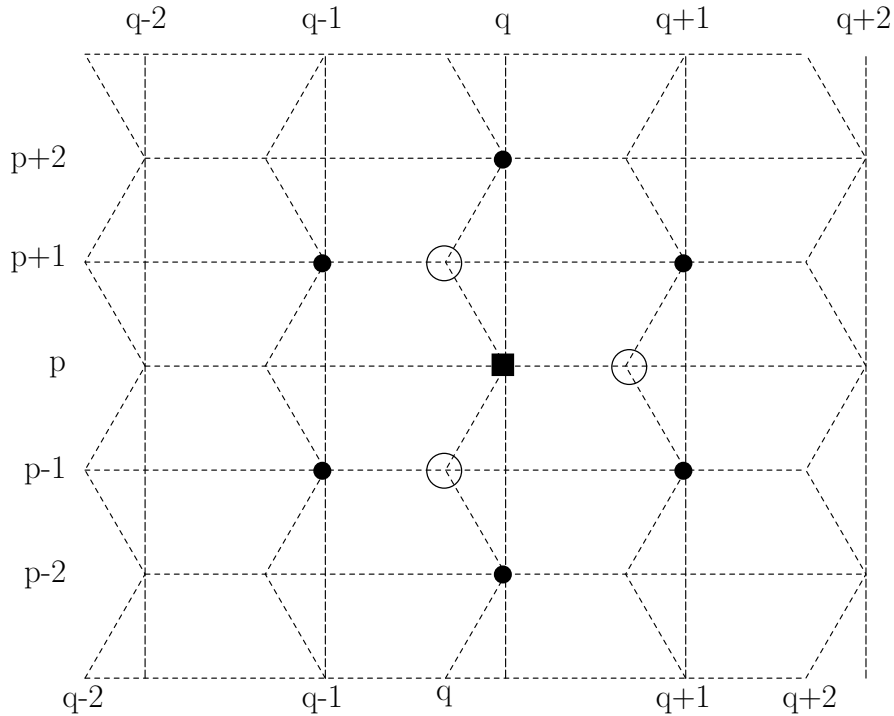}
\caption{Honeycomb lattice showing a site
at $(p,q)$ (filled box) surrounded by nearest neighbours on a honeycomb lattice
(large circles) and nearest neighbours on a triangular lattice
(small filled circles).
The $q$ labels at the bottom of the figure refer to
the honeycomb lattice and those at the top refer to the triangular lattice.}
\end{figure}
This relationship between random walks on the honeycomb lattice and random walks
on the triangular lattice is illustrated in Figure 3.
In this figure the site at $(p,q)$ is shown as a $\vdash$ site.
The nearest neighbour triangular lattice sites to the site at $(p,q)$
are of the same $\vdash$ symmetry type and co-ordinate labels for the honeycomb
lattice and the triangular lattice match at all $\vdash$ sites.
A similar matching occurs
at $\dashv$ sites (as can be seen by inverting Figure 3).
The nearest neighbours on the triangular lattice are next nearest
neighbours on the honeycomb lattice and the probability of a random walk from
$(p,q)$ to one of these sites is the same for  random walks on 
 the triangular
lattice and the honeycomb lattice.

The triangular lattice equation fails for $F(p,n)$ due
to the absorbing boundary condition $\hat F(p,n+1)=0$ and similarly the
triangular lattice equation fails for $\hat F(p,1)$ due to the absorbing
 boundary condition $F(p,0)=0$.
To circumvent these boundary problems we use the homogeneous
triangular lattice equation
solutions
for $F(p,q)$ in the region $q\le b$ (region I) and we use the
homogeneous triangular lattice
equation solutions for $\hat F(p,q)$ in the region $q\ge b+1$
(region II).
Thus we have homogeneous solutions of the form
\begin{eqnarray}
F_I(p,q)&=&cq
+d q\cos[\pi(p-a)]\cos[\pi(q-b)]\nonumber\\
& &+{\sum_{k=1}^m}^\prime (a_k e^{i\alpha_k p}+b_k e^{-i\alpha_k p})\sinh(\gamma_k q) ,\label{Hh1}\\
\hat F_{II}(p,q)&=&\hat c(q-n-1)\nonumber\\
& &+\hat d (q-n-1)\cos[\pi(p-a)]\cos[\pi(q-b)]\nonumber\\
& &+{\sum_{k=1}^m}^\prime (\hat a_k e^{i\alpha_k p}+\hat b_k e^{-i\alpha_k p})\sinh[\gamma_k(q-n-1)].\label{Hh2}
\end{eqnarray}
These homogeneous solutions satisfy the periodic boundary conditions
as well as the absorbing boundary conditions $\hat F_{II}(p,n+1)=0$ and
$F_{I}(p,0)=0$.

We now consider a single point source
at a type $\vdash$ symmetry site.
Thus we have the inhomogeneous problem
\begin{eqnarray}
F_{I}(p,b)&=&\delta_{p,a}+\frac{1}{2+\eta}\left[\eta\hat F_{II}(p,b+1)\right.
\nonumber\\
& &\left. + \hat F_{I}(p+1,b)+\hat F_{I}(p-1,b)\right]\label{HI1}\\
\hat F_{II}(p,b+1)&=&\frac{1}{2+\eta}\left[\eta F_{I}(p,b) +  F_{II}(p+1,b+1)\right.\nonumber\\
& &\left.+ F_{II}(p-1,b+1)\right].\label{HI2}
\end{eqnarray}
At this stage we do not have general solutions for $\hat F_{I}(p,q)$ and
$F_{II}(p,q)$ and so we use Eqs.(\ref{H1}),(\ref{H2}) to write the
inhomogeneous problem in the form
\begin{eqnarray}
F_I(p,b)&=&\left[ (2+\eta)^2\delta_{p,a}+(2+\eta)\eta \hat F_{II}(p,b+1)\right.\nonumber\\
&+&\eta F_{I}(p+1,b-1)+F_{I}(p+2,b)\nonumber\\
&+&\left.\eta F_{I}(p-1,b-1)+F_{I}(p-2,b)\right]\nonumber\\
&/&\left[(2+\eta)^2-2\right]\label{I1}
\end{eqnarray}
\begin{eqnarray}
\hat F_{II}(p,b+1)&=&\left[ (2+\eta)\eta F_{I}(p,b)\right.\nonumber\\
&+&\eta \hat F_{II}(p+1,b+2)+\hat F_{II}(p+2,b+1)\nonumber\\
&+&\eta \left.\hat F_{II}(p-1,b+2)+\hat F_{II}(p-2,b+1)\right]\nonumber\\
&/&\left[(2+\eta)^2-2\right].\label{I2}
\end{eqnarray}
The unknown constants $c,d,a_k,b_k,\hat c, \hat d, \hat a_k, \hat b_k$ can
now be obtained by substituting
the homogeneous solutions, Eqs.(\ref{Hh1}),(\ref{Hh2})
and the Kronecker delta identity, Eq.(\ref{Id2}),
into the inhomogeneous equations Eqs.(\ref{I1}),(\ref{I2})
and equating linearly independent functions of $p$.
The algebraic manipulations are simplified using the
identities in Eqs.(\ref{A2}),(\ref{A3}).

The resulting expressions for the expectation values are
\begin{widetext}
\begin{eqnarray}
F_I(p,q)&=&\frac{(2+\eta)^2}{2\eta(m+1)}
\left(1-\frac{(\eta+2)b}{(\eta+2)n+2}\right)q
(1+\cos[\pi(p-a)]\cos[\pi(q-b)])\nonumber\\
&+&\frac{(2+\eta)^2}{2\eta(m+1)}{\sum_{k=1}^m}^\prime
\left\{\cos[\alpha_k(p-a)]\sinh(\gamma_k q)\right.\nonumber\\
&\times&\left.\left[(\eta+4\cosh(\gamma_k)\cos(\alpha_k))\sinh[\gamma_k(b-n)]
-2\cos\alpha_k\sinh[\gamma_k(b+1-n)]\right]\right\}\nonumber\\
&/&\left\{
(\eta\cosh\gamma_k\cos\alpha_k-1-\eta)(\cosh[\gamma_k(n-2)]
-\cosh(\gamma_k n))\right.\nonumber\\
&-&
\left.(\eta+4\cosh\gamma_k\cos\alpha_k)\cos\alpha_k\sinh\gamma_k\sinh(\gamma_k n)\right\}
\label{HS1}
\end{eqnarray}
\begin{eqnarray}
\hat F_{II}(p,q)&=&\frac{(2+\eta)^3b(n+1-q)}{2(\eta n+2n+2)\eta(m+1)}(1-\cos[\pi(p-a)]\cos[\pi(q-b)])\nonumber\\
&+&\frac{(2+\eta)^3}{2\eta(m+1)}{\sum_{k=1}^m}^\prime
\left\{\cos[\alpha_k(p-a)]\sinh[\gamma_k (q-n-1)]\sinh(\gamma_k b)\right\}
\nonumber\\
&/&\left\{
(\eta\cosh\gamma_k\cos\alpha_k-1-\eta)(\cosh[\gamma_k(n-2)]
-\cosh(\gamma_k n))\right.\nonumber\\
&-&
\left.(\eta+4\cosh\gamma_k\cos\alpha_k)\cos\alpha_k\sinh\gamma_k\sinh(\gamma_k n)\right\}
\label{HS2}
\end{eqnarray}
The expectation values at type $\dashv$ sites in region I and $\vdash$ sites 
in region II can now be obtained
by substituting the solutions from Eq.(\ref{HS1}) into  Eq.(\ref{H2}) and
the solutions from Eq.(\ref{HS2}) into Eq.(\ref{H1}) respectively.
The results are
\begin{eqnarray}
\hat F_{I}(p,q)&=&\frac{(2+\eta)}{2\eta(m+1)}
\left(1-\frac{(\eta+2)b}{(\eta+2)n+2}\right)((\eta+2)q-\eta)
(1-\cos[\pi(p-a)]\cos[\pi(q-b)])\nonumber\\
&+&\frac{(2+\eta)}{2\eta(m+1)}{\sum_{k=1}^m}^\prime
\left\{\cos[\alpha_k(p-a)][\eta\sinh[\gamma_k (q-1)]+2\cos(\alpha_k)\sinh(\gamma_k q)]\right.\nonumber\\
&\times&\left.\left[\left(\eta+4\cosh(\gamma_k)\cos(\alpha_k)\right)\sinh[\gamma_k(b-n)]
-2\cos\alpha_k\sinh[\gamma_k(b+1-n)]\right]\right\}\nonumber\\
&/&\left\{
(\eta\cosh\gamma_k\cos\alpha_k-1-\eta)(\cosh[\gamma_k(n-2)]
-\cosh(\gamma_k n))\right.\nonumber\\
&-&
\left.(\eta+4\cosh\gamma_k\cos\alpha_k)\cos\alpha_k\sinh\gamma_k\sinh(\gamma_k n)\right\}
\label{HS3}
\end{eqnarray}
\begin{eqnarray}
F_{II}(p,q)&=&\frac{(2+\eta)^2b[(\eta+2)(n-q)+2]}{2((\eta+2) n+2)\eta(m+1)}(1+\cos[\pi(p-a)]\cos[\pi(q-b)])\nonumber\\
&+&\frac{(2+\eta)^2}{2\eta(m+1)}{\sum_{k=1}^m}^\prime
\left\{\cos[\alpha_k(p-a)]
\sinh(\gamma_k b)\nonumber\right.\\
&\times&
\left.\left(\eta\sinh[\gamma_k(q-n)]+2\cos(\alpha_k)\sinh[\gamma_k(q-n-1)]\right)
\right\}
\nonumber\\
&/&\left\{
(\eta\cosh\gamma_k\cos\alpha_k-1-\eta)(\cosh[\gamma_k(n-2)]
-\cosh(\gamma_k n))\right.\nonumber\\
&-&
\left.(\eta+4\cosh\gamma_k\cos\alpha_k)\cos\alpha_k\sinh\gamma_k\sinh(\gamma_k n)\right\}
\label{HS4}
\end{eqnarray}

\end{widetext}
The absorption probabilities are
defined by
\begin{eqnarray}
\hat G(p,0)&=&0,\\
G(p,0)&=&\frac{\eta}{2+\eta} \hat F_I(p,1),\\
\hat G(p,n+1)&=&\frac{\eta}{2+\eta} F_{II}(p,n),\\
G(p,n+1)&=&0.
\end{eqnarray}
The total absorption probabilities at the ends of the tubes are thus
$$
\sum_{p=0}^m G(p,0)=1-\frac{(\eta+2)b}{(\eta+2)n+2}
$$
and
$$
\sum_{p=0}^m G(p,n+1)=\frac{(\eta+2)b}{(\eta+2)n+2}.
$$
Note that the absorption probablilities in this case are functions of the 
bias parameter $\eta$. The square lattice tube and the triangular lattice 
tube are symmetric with respect to left/right walks along the axial direction 
and thus the probabilities for absorption at the ends depend only on the 
initial distance from the ends at which particles are released. The honeycomb 
lattice is not symmetric with respect to left/right walks along the axial 
direction. As a consequence the absorption probabilities at the ends of the 
tube depend on both the initial distance from the ends (which also determines 
the symmetry type, $\vdash$ or $\dashv$, of the initial lattice site) and the 
axial bias parameter.

\section{Example and Discussion}
In this paper we have derived exact 
formulae for the expectations that a random walk starting
at a lattice point $(a,b)$ will visit a lattice site $(p,q)$
on a lattice tube with absorbing lattice sites on the ends.
The formulae
for square lattice tubes, Eqs.(\ref{SS1}),(\ref{SS2}),
triangular lattice tubes, Eqs.(\ref{T1}),(\ref{T2}), and
 honeycomb lattice cubes,
Eqs.(\ref{HS1}),(\ref{HS2}),(\ref{HS3}),(\ref{HS4}) allow us to readily compute the expectation values
for tubes of any specified size and arbitrary starting points.
Moreover each of these solutions contains an adjustable parameter
$\eta$ that can be adjusted away from unity to model different
random walk probabilities in the cyclic direction around the tube
compared with the axial direction along the tube.

As an example we consider the case of a honeycomb lattice tube
with $m=17, n=29$ and three values of $\eta$; i) $\eta=1$,
ii) $\eta=1/100$, 
and iii) $\eta=100$. We have taken the source to be centrally located
at $a=9, b=15$ in each case.
The expectation values at each of the lattice co-ordinates have been plotted
in Figure 4.
As might be anticipated, 
for small values of the axial bias parameter, $\eta\ll 1$,
diffusion along the tube axis is very slow;
the random walk cycles around the tube 
many times (several hundred times for $\eta=1/100$) before finally 
being absorbed at
one of the open ends. For large values of the
axial bias parameter, $\eta\gg 1$,
we might first anticipate rapid
diffusion along the tube axis however the tower
like plot in Figure 4(c) reveals that this is not the case. 
The random walk becomes trapped locally near the source as it moves back
 and forward between the source site and the nearest neighbour to the source.
This effect is indeed a simple consequence of the honeycomb lattice
geometry which has only one  nearest neighbour along the axis direction.

\begin{figure}[h!]
\vspace{200mm}
\includegraphics{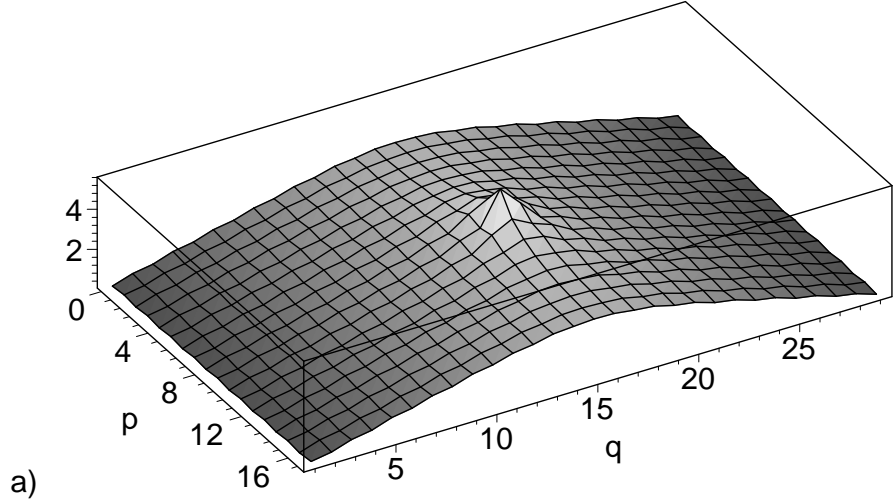}
\includegraphics{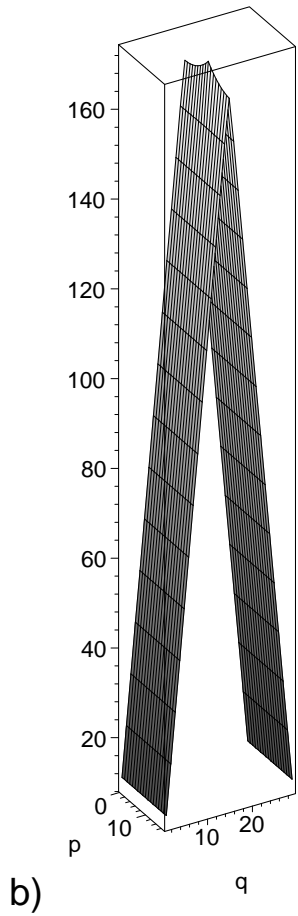}
\includegraphics{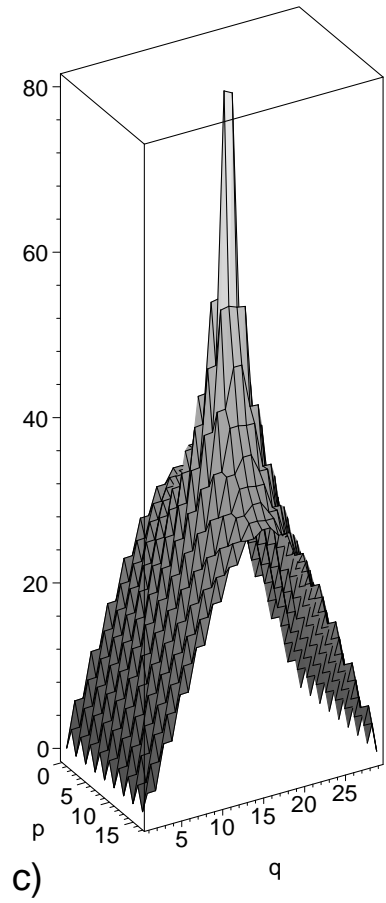}
\caption{Expectation values for a random walk on a honeycomb tube
for three different values of the axial bias; a) $\eta=1$,
b) $\eta=0.01$, c) $\eta=100$.}
\end{figure}

A further interesting calculation is the steady state profile
 for expectation values along the lattice
tube after summation over $p$. The profile is piecewise linear
with a linear increase from $q=1$ up to $q=b$ followed
 by a linear decrease from $q=b$ to $q=n$.
The slope of the linear portions is dependent on the parameter $\eta$.
Explicit expressions for this  slope as a function of $\eta$
can be readily evaluated from the formulae for the expectation
values given in Eqs (54)-(57).
For example for  $q\le b$ we have
\begin{eqnarray*}
E_I(q)&=&\sum_{p=0}^m F_I(p,q)+\hat F_I (p,q)\\
&=&\frac{(\eta+2)^2}{\eta}\left(1-\frac{(n+2)b}{(\eta+2)n+2}\right) q
- \frac{(\eta+2)}{2}\left(1-\frac{(\eta+2)b}{(\eta+2)n+2}\right),
\end{eqnarray*}
It is clear from this equation that the slope
diverges as $\eta\to 0$ and as $\eta\to\infty$. 
For the example considered here with $n=29$ and $b=15$ we have
$$
E_I(q)=\frac{(7\eta+15)(\eta+2)(2(\eta+2)q-\eta)}{\eta(60+29\eta)}
$$
and the slope is a minimum at $\eta\approx 2.035$.

The geometry of the honeycomb lattice tube
that we have considered in this paper is  equivalent to that of a single walled
 zig-zag carbon nanotube with open ends. The above example corresponds to
the (9,0) nanotube in the standard notation \cite{O98}.
The bias parameter could thus be tuned to model the
effects of strain, such as surface curvature,
 on diffusion of adatoms along the carbon-carbon bonds
 on zig-zag carbon nano-tubes.
An interesting result in this connection
(as shown in Figure 4(c) above) is that
a random walk could be localized for a period of time
by applying a uniform strain which favours diffusion
in the direction of all bonds
aligned with the tube axis.

\appendix*
\section{}
The following identites have proven useful for
deriving the results in this paper
\begin{eqnarray}
&&\sinh(\gamma(b-1))\sinh(\gamma(b-n-1))+\sinh(\gamma b)\sinh(\gamma(b-n))\nonumber\\
&&=\sinh(\gamma)\sinh(\gamma(n+1))+
2\cosh(\gamma)\sinh(\gamma b)\sinh(\gamma(b-n-1))\label{A1}
\end{eqnarray}
\begin{eqnarray}
&&\sinh(\gamma(b+1-n))\sinh(\gamma b)+\sinh(\gamma (b-1))\sinh(\gamma(b-n))\nonumber\\
&&=\sinh(\gamma)\sinh(\gamma n )+2\cosh(\gamma)\sinh(\gamma b)\sinh(\gamma(b-n))\label{A2}\end{eqnarray}
\begin{eqnarray}
&&\sinh(\gamma(b-n))\sinh(\gamma b)+\sinh(\gamma (b-1))\sinh(\gamma(b+1-n))\nonumber\\
&&=\left(\cosh(\gamma(n-2))-\cosh(\gamma n)\right)/2\label{A3}
\end{eqnarray}


\begin{references}
\bibitem{ML79}
E.W. Montroll and B.J. West,
On an enriched collection of stochastic processes,
in \textit{Fluctuation Phenomena,} edited by
E.W. Montroll and J.L. Lebowitz
(Elsevier Science Publishers B.V., Amsterdam, 1979)

\bibitem{D53}
R.J. Duffin,
Discrete potential theory.
Duke Math. J.
\textbf{20}, 233 (1953).

\bibitem{DS84}
P.G. Doyle and J.L. Snell,
\textit{Random Walks and Electrical Networks}
(Mathematical Association of America, Washington D.C.,
(1984).

\bibitem{BCMO99}
O. Benichou, A.M. Cazabat, M. Moreau, and G. Oshanin,
Directed random walk in adsorbed monolayer.
Physica A
\textbf{272}, 56 (1999).

\bibitem{HPAB99}
R. Holyst, D. Plewczynski, A. Aksimentiev and
K. Burdzy,
Diffusion on curved periodic surfaces,
Phys. Rev. E
\textbf{60}, 302 (1999).

\bibitem{C28}
R. Courant, K. Friedrichs, and H. Lewy, 
Uber die partiellen Differenzengleichungen der mathematischen Physik.
Math. Ann. \textbf{100}, 32 (1928).

\bibitem{MW40}
W.H. McCrea and  F.J.W. Whipple, 
Random paths in two and three dimensions.
Proc. Roy. Soc. Edinburgh.
\textbf{60}, 281 (1940).

\bibitem{BH02c}
M.T. Batchelor and B.I. Henry,
Gene Stanley, the $n$-vector model and random walks with
absorbing boundaries,
Physica A. \textbf{314}, 77 (2002).

\bibitem{BH02}
M.T. Batchelor and B.I. Henry,
Exact solution for random walks on the triangular lattice with
absorbing boundaries,
J. Phys. A: Math. Gen. \textbf{35}, 5951 (2002).

\bibitem{KM63}
E.M. Keberle and G.L. Montet, 
Explicit solutions of partial difference equations and random
paths on plane nets.
J. Math. Anal. Appl.
\textbf{6},
 1 (1963).

\bibitem{M94}
J.W. Miller,
A matrix equation approach to solving recurrence relations in
two-dimensional random walks.
J. Appl. Prob.
\textbf{31}, 646 (1994).

\bibitem{FZ01}
M. Ferraro and L. Zaninetti,
Number of times a site is visited in two-dimensional random
walks.
Phys. Rev. E,
\textbf{64}, 056107 (2001). 


\bibitem{F66}
W. Feller,
\textit{An Introduction to Probability Theory and Its Applications}
(Wiley, New York, 1966). p. 330.


\bibitem{LJLLWXLTZZ02}
J-L. Li, J-F. Jia, X-J. Liang, X. Liu, J-Z. Wang, Q-K Xue,
Z-Q. Li, J.S. Tse, Z. Zhang and S.B. Zhang,
Spontaneous assembly of perfectly ordreed identical-size nanocluster
arrays,
Phys. Rev. Letts. \textbf{88}, 0066101 (2002)

\bibitem{O98} T.W. Odom, J.-L. Huang, P. Kim and C.M. Lieber,
Atomic structure and electronic properties
of single-walled carbon nanotubes,
Nature \textbf{391}(6662), 62 (1998)

\bibitem{KO01} N. Kitamura and A. Oshiyama,
Open edge growth mechanisms of single wall carbon nanotubes,
J. Phys. Soc. Japan. \textbf{70}, 1995 (2001)

\bibitem{LSK02} O.A. Louchev, Y. Sato and H. Kanda,
Morphologiocal stabilization, destabilization, and open-end closure
during carbon nanotube growth mediated by surface diffusion,
Phys. Rev. E \textbf{66}, 011601 (2002)

\bibitem{C00} N. Coftas,
Random walks on carbon nanotubes and quasicrystals,
J. Phys. A. \textbf{33}, 2917 (2000)

\bibitem{SG01}
D.J. Shu and X.G. Gong,
Curvature effect on surface diffusion: The nanotube,
J. Chem. Phys. \textbf{114}, 10922 (2001)
\end{references}
\end{document}